# High-performance near- and mid-infrared crystalline coatings

Garrett D. Cole,[1,2*] Wei Zhang,[3] Bryce J. Bjork,[3] David Follman,[1] Paula Heu,[1] Christoph Deutsch,[2] Lindsay Sonderhouse,[3] John Robinson,[3] Chris Franz,[4] Alexei Alexandrovski,[4] Mark Notcutt,[5] Oliver H. Heckl,[3] Jun Ye,[3] and Markus Aspelmeyer[6]

[1]*Crystalline Mirror Solutions LLC, Santa Barbara, California 93101, USA*
[2]*Crystalline Mirror Solutions GmbH, 1010 Vienna, Austria*
[3]*JILA, National Institute of Standards and Technology and University of Colorado, Boulder, Colorado 80309, USA*
[4]*Stanford Photo-Thermal Solutions, Pahoa, Hawaii 96778, USA*
[5]*Stable Laser Systems, Boulder, Colorado 80301, USA*
[6]*Vienna Center for Quantum Science and Technology (VCQ), Faculty of Physics, University of Vienna, A-1090 Vienna, Austria*
*Corresponding author: g.cole@crystallinemirrors.com

Substrate-transferred crystalline coatings have recently emerged as a groundbreaking new concept in optical interference coatings. Building upon our initial demonstration of this technology, we have now realized significant improvements in the limiting optical performance of these novel single-crystal GaAs/Al$_x$Ga$_{1-x}$As multilayers. In the near-infrared (NIR), for coating center wavelengths spanning 1064 to 1560 nm, we have reduced the excess optical losses (scatter + absorption) to levels as low as 3 parts per million, enabling the realization of a cavity finesse exceeding $3 \times 10^5$ at the telecom-relevant wavelength range near 1550 nm. Moreover, we demonstrate the direct measurement of sub-ppm optical absorption at 1064 nm. Concurrently, we investigate the mid-IR (MIR) properties of these coatings and observe exceptional performance for first attempts in this important wavelength region. Specifically, we verify excess losses at the hundred ppm level for wavelengths of 3300 and 3700 nm. Taken together, our NIR optical losses are now fully competitive with ion beam sputtered multilayer coatings, while our first prototype MIR optics have already reached state-of-the-art performance levels for reflectors covering this portion of the fingerprint region for optical gas sensing. Mirrors fabricated with our crystalline coating technique exhibit the lowest mechanical loss, and thus the lowest Brownian noise, the highest thermal conductivity, and, potentially, the widest spectral coverage of any "supermirror" technology in a single material platform. Looking ahead, we see a bright future for crystalline coatings in applications requiring the ultimate levels of optical, thermal, and optomechanical performance.

## 1. Introduction

Highly reflective optical interference coatings are indispensable tools for modern scientific and industrial efforts. Systems with ultralow optical losses, namely parts per million (ppm) levels of scatter and absorption, were originally developed for the construction of ring-laser gyroscopes by Litton Systems in the late 1970s [1]. Stemming from this breakthrough, ion-beam sputtering (IBS) is now firmly established as the gold standard process technology for generating ultralow-loss reflectors in the visible and near infrared (NIR) [2]. Typically, such multilayers consist of alternating layers of amorphous metal-oxides, most commonly high index Ta$_2$O$_5$ (tantala) and low index SiO$_2$ (silica) thin films, finding application in narrow-linewidth laser systems for optical atomic clocks [3, 4], gravitational wave detectors [5, 6], cavity QED [7], and tests of fundamental physics [8]. Though exhibiting phenomenal optical properties, limitations of these amorphous coatings include excess Brownian noise [9], negatively impacting the ultimate performance of precision optical interferometers, poor thermal conductivity (typically below 1 Wm$^{-1}$K$^{-1}$), as well as significant levels of optical absorption for wavelengths beyond 2 μm, excluding operation of these reflectors in the mid-infrared (MIR). The latter limitation means that the highest performing metal oxide structures, while exhibiting phenomenal performance in the visible and NIR, cannot operate with low losses in this important wavelength region requiring a switch to amorphous II-VI, group IV, or IV-VI compounds.

In this Article we present important advancements in the development of an entirely new class of ultra-low loss optical interference coatings. These mirrors, based on substrate-transferred single-crystal semiconductor heterostructures (henceforth referred to as "crystalline coatings") can now achieve optical performance rivaling that of IBS multilayers in both the NIR and MIR. Moreover, mirrors fabricated via this technique exhibit vastly reduced Brownian noise, the highest thermal conductivity (>30 Wm$^{-1}$K$^{-1}$ [10] compared to <1 Wm$^{-1}$K$^{-1}$ for SiO$_2$/Ta$_2$O$_5$), and, potentially, the widest spectral coverage of any supermirror technology, owing to state-of-the art levels of scatter and absorption losses in both the near and mid IR. Here we present an in-depth investigation of the optical performance of these novel coatings, demonstrating NIR (1064-1550 nm) reflectors with excess optical losses (scatter + absorption) as low as 3 ppm in the best optics, with independent measurements of the optical absorption yielding values at or below 1 ppm at 1064 nm. Pushing into the MIR, in this case for mirrors operating in the range from 3 to 4 μm, we measure excess losses of 159 ppm and 242 ppm for center wavelengths of 3.3 and 3.7 μm respectively. These low loss levels enable the demonstration of a cavity with a finesse exceeding $1 \times 10^4$ at 3.3 μm, while simultaneously exhibiting a cavity reflection contrast of 71%. At 3.7 μm, we realize a cavity finesse of approximately $5 \times 10^3$ with a significantly higher transmission exceeding 500 ppm, yielding a cavity contrast of 90%. Our first attempts at fabricating MIR mirrors have already resulted in reflectors with optical losses on par with the best coatings present on the commercial market, yielding optical enhancement cavities with state-of-the-art values of cavity finesse and resonance efficiency.

## 2. Near-infrared crystalline coatings

Coating Brownian noise, driven by excess mechanical dissipation in high-reflectivity IBS optical coatings imposes a severe limit on the performance of state-of-the-art precision measurement systems, such as stabilized lasers for optical atomic clocks [11, 12] and interferometric gravitational wave detectors [5, 6]. As a consequence, a concerted effort has been focused on the identification of high-reflectivity multilayers capable of simultaneously achieving minimal mechanical dissipation. The first indication of the potential for improved Brownian noise performance in crystalline multilayers was originally revealed in the pursuit of fundamental cavity optomechanics research [13-16]. Following this initial work, we transitioned from micrometer-scale resonators to cm-scale optics, developing a suitable high-yield bonding process that relies on advanced semiconductor microfabrication techniques and enables the integration of low-loss epitaxial multilayers with standard super-polished substrates [17].

This novel substrate-transfer coating procedure entails separating the epitaxial multilayer from its original growth wafer and directly bonding it—without the use of adhesives or intermediate films—to a desired host substrate (Fig. 1a). With this technique, the bonded mirror assembly initially begins as two separate components, a GaAs wafer capped with an epitaxial multilayer and a super-polished optical substrate with a standard backside antireflection coating. The single-crystal multilayer is grown using molecular beam epitaxy (MBE) on a 150-mm diameter semi-insulating GaAs wafer and is comprised of alternating quarter-wave optical thickness GaAs for the high-index layers and $Al_{0.92}Ga_{0.08}As$ for the low-index films. The lateral geometry of the mirror disc is defined by optical lithography, followed by chemical etching to extrude the coating shape through the epitaxial multilayer. Chemo-mechanical substrate removal using lapping, followed by selective wet chemical etching, is then used to strip the GaAs growth template. Next, a thick AlGaAs etch stop layer, incorporated beneath the Bragg stack is removed and the mirror surface is cleaned of any potential debris. Finally, the crystalline mirror disc and silica substrate are pressed into contact, resulting in a spontaneous van der Waals bond. To strengthen the interface and minimize potential frictional losses, a post-bond anneal completes the fabrication procedure.

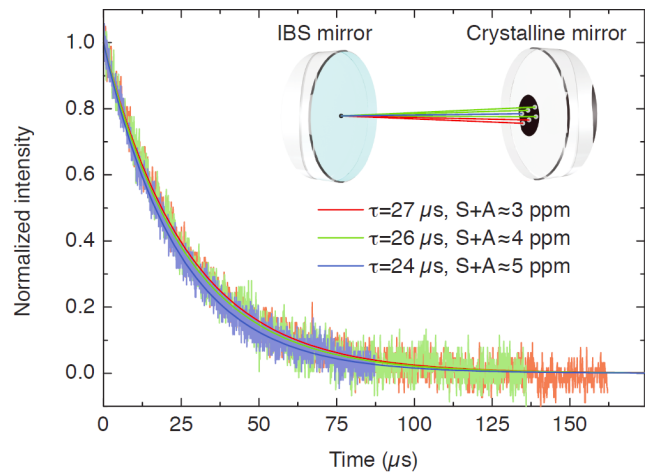

*Fig. 2. Position-dependent ringdown of a 1550 nm crystalline coating with a nominal transmission of 10 ppm. As indicated in the inset, six discrete positions in a 3×2 configuration with a roughly 1-mm spacing are probed in a 75-mm long cavity employing one fused silica substrate with a crystalline coating and one reference IBS-coated mirror having a total optical loss of 6 ppm. We observe a weak position dependence in the excess losses (scatter + absorption, S+A), with all points ≤ 5 ppm for the optimized crystalline coating.*

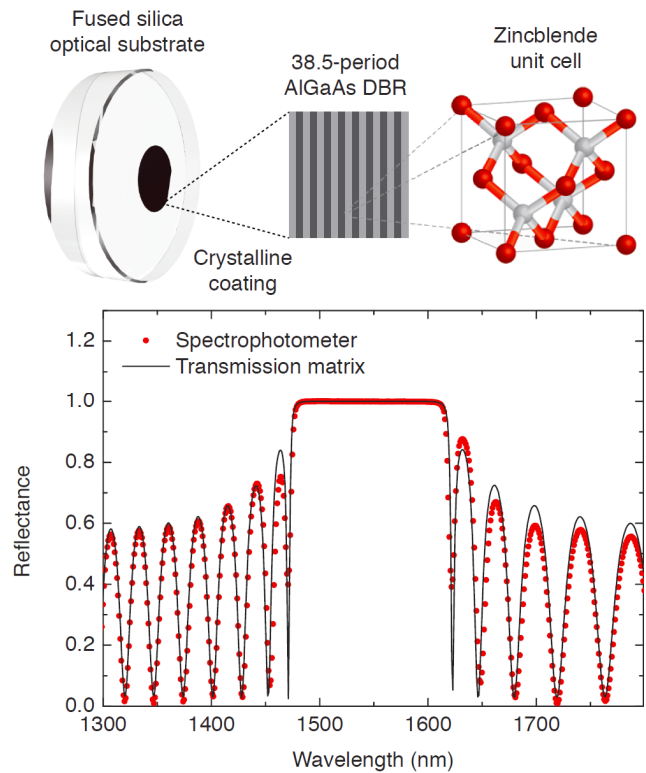

*Fig. 1. Details of our NIR crystalline coatings. Top: Schematic of a crystalline supermirror consisting of a super-polished fused silica substrate with a substrate-transferred NIR crystalline coating. The first inset shows the repeating and alternating high index GaAs and low index $Al_{0.92}Ga_{0.08}As$ layers, while the second shows the zincblende unit cell of which the coating is comprised. In this Article we present results for NIR multilayers with both 38.5 periods (shown here), with a nominal transmission of 10 ppm (yielding a finesse of $2 \times 10^5$) and 41.5 periods, reducing the transmission to 5 ppm and increasing the cavity finesse to $3 \times 10^5$ at 1550 nm when transferred to fused silica substrates. Bottom: Measured reflectance spectrum for a 38.5-period quarter-wave epitaxial multilayer with a center wavelength of 1550 nm following removal from the original GaAs growth wafer and bonding to a super-polished fused silica substrate with a 1-m ROC.*

In 2013 we experimentally verified that such substrate-transferred crystalline coatings are able to maintain their high mechanical quality factors post-bonding, with an exceptionally low mechanical loss angle <$4 \times 10^{-5}$ extracted from measurements of cavity-stabilized laser noise performance, representing a tenfold reduction when compared with the best dielectric multilayers at room temperature [17]. Since that time, low-loss crystalline coatings have been employed for the demonstration of a large-area NIR ring-laser gyroscope [18] and have also shown the potential for the minimization of thermo-optic noise by inducing coherent cancellation of the thermoelastic and thermorefractive noise components in a specially designed multilayer [19]. Moreover, these coatings have been successfully applied to a variety of substrate materials including fused silica, silicon, sapphire, SiC, diamond, YVO, and YAG, among others, with recent advancements enabling the transfer of crystalline multilayers onto substrates with radii of curvature as tight as 100 mm (for a 5-mm diameter coating). With respect to the maximum achievable coating area, with >30-cm diameter optics required for the test masses in interferometric gravitational wave detectors [5, 6], the process can immediately be extended to a maximum coating diameter of 20 cm for less severely curved surfaces, currently limited by the maximum available GaAs substrate size. Note that this limitation may be overcome if suitable large-area substrates can be developed, namely GaAs wafers with diameters >20 cm, or through the use of heteroepitaxial growth on Si via relaxed SiGe buffer layers [20], with the potential for expanding the coating diameter to the current maximum Si wafer size of 450 mm.

Though these initial demonstrations prove that we had developed a robust manufacturing process capable of generating high-reflectivity coatings with exceptional thermal noise performance, our early prototype mirrors suffered from rather high levels of optical losses, with scatter and absorption totaling 20 ppm at 1064 nm [17]. Over the ensuing two years, a significant effort has been undertaken to both understand and further improve the optical performance of these novel coatings, particularly the position dependence of the optical losses, through optimization of the crystal growth and substrate-transfer processes. With a focused effort on minimizing the background impurity level of the constituent films during the epitaxial growth process, we have achieved optical absorption values below 1 ppm in the NIR, with direct loss measurements undertaken via photothermal common path interferometry (PCI) [21], as well as cavity ringdown at a number of wavelengths in the span of 1064 nm to 1560 nm. Most recently, by reducing the impact of embedded defects via a modified substrate-transfer process, we have also largely suppressed optical scatter losses, reaching limiting levels of ~3 ppm in the same wavelength range.

In our newly devised microfabrication technique, we reduce the impact of surface defects (commonly referred to as "oval defects" and originating from spitting of group III metal droplets from Knudsen-type effusion cells [22]) by burying them at the bond interface. In this case, with no means of planarization during the deposition process, each defect incorporated in the course of the crystal growth process results in a small perturbation at the top surface of the epitaxial stack, with the exact geometry depending on the "seed" profile as well as the depth within the multilayer. Typically, such buried defects generate nm-scale hillocks roughly 1-10 μm in lateral size, that do not perturb the direct bonding process, though in extreme cases, these defects may be more than 1 μm tall. As we have access to the front and backside of the multilayer in our substrate-transfer procedure, we minimize the impact of these defects on the overall optical loss by "flipping" the coating before bonding, leaving the initial backside exposed to the optical field, with the top surface of the crystal in contact with the final optical substrate. What was originally the back surface of the epitaxial multilayer shows a significantly reduced defect density and essentially replicates the substrate surface quality (with a ~1 Å RMS roughness surface realized after the etch stop removal process), with minimal defects incorporated from the buffer and etch stop layer deposition steps. Using this modified bonding process, we have now realized crystalline coatings with low loss levels and, even more importantly, improved uniformity in terms of the surface quality and thus optical scattering losses.

Exploiting this updated substrate-transfer process, we fabricate crystalline-coated cavity end mirrors with optical scatter and absorption losses as low as 3 ppm, with direct measurements near 1064 nm, 1400 nm, and 1550 nm. Here we highlight detailed measurements of high-finesse 1550 nm optics. The first example consists of a 38.5-period (9.50-μm thick) $GaAs/Al_{0.92}Ga_{0.08}As$ multilayer transferred to super-polished fused silica, resulting in 10 ppm transmission at a center wavelength of 1550 nm. For the experiment, a resonant cavity is built using mirrors held in standard mirror mounts, including one substrate coated with the crystalline multilayer under test and the second consisting of a reference IBS-deposited dielectric multilayer; the latter reference mirror having been previously characterized in terms of its optical loss and transmission level. Employing this hybrid test cavity, the sum of the loss and transmission of the crystalline coating can thus be extracted from the cavity ringdown time.

In order to explore the uniformity of the crystalline coating, the mirror under test is laterally displaced with respect to the cavity optical axis. Position-dependent optical ringdown measurements, as shown in Figure 2, yield excess loss levels of at worst 5 ppm. For this exemplary mirror, six roughly equally spaced positions over an 8-mm diameter coating were probed, revealing minimal position dependence of the optical loss. This mirror yields a high uniformity, with excess losses extracted from ringdown measurements in transmission falling between 3 and 5 ppm. With a nominal transmission of 10 ppm, two identical mirrors of this type incorporated in a rigid cavity yield a cavity finesse of just over $2 \times 10^5$ with a contrast of 30% in reflection.

Increasing the number of layer pairs to 41.5 (total thickness of 10.2 μm), the transmission is further reduced to 5 ppm and the maximum achievable finesse is increased. Using two of these low transmission mirrors to construct a 240-mm long cavity, we have now experimentally verified a maximum cavity finesse of $3 \times 10^5$ near 1550 nm, as shown in Figure 3. Figure 3a displays finesse measurements as a function of beam position on the mirror. The measured finesse in this case shows a stronger position dependence than the example above, with two points yielding finesse values between $1-2 \times 10^5$. This result indicates that there is still the potential for a variation in optical losses, which currently requires probing for mirrors and mirror positions that ultimately meet the desired levels of performance, particularly at these very low levels of excess loss. With further development, we are confident that the manufacturing process will allow for the routine realization of such high finesse values. Figure 3b shows the 77 μs ringdown time corresponding to the position of highest finesse, in this case $3 \times 10^5$ at 1544 nm. These results represent a significant enhancement in optical quality over our initial demonstration from 2013 [17], enabling crystalline coatings to reach a level of optical performance on par with IBS multilayers. It is also important to note that, based on previous measurements of the absorption coefficients of GaAs and AlGaAs, these systems do not exhibit any narrow absorption lines. Thus, we are confident that this performance level can be maintained for any center wavelength over the range of ~1000-2000 nm and eventually for much longer wavelengths, potentially even out to the maximum wavelength of transparency at ~10 μm for GaAs.

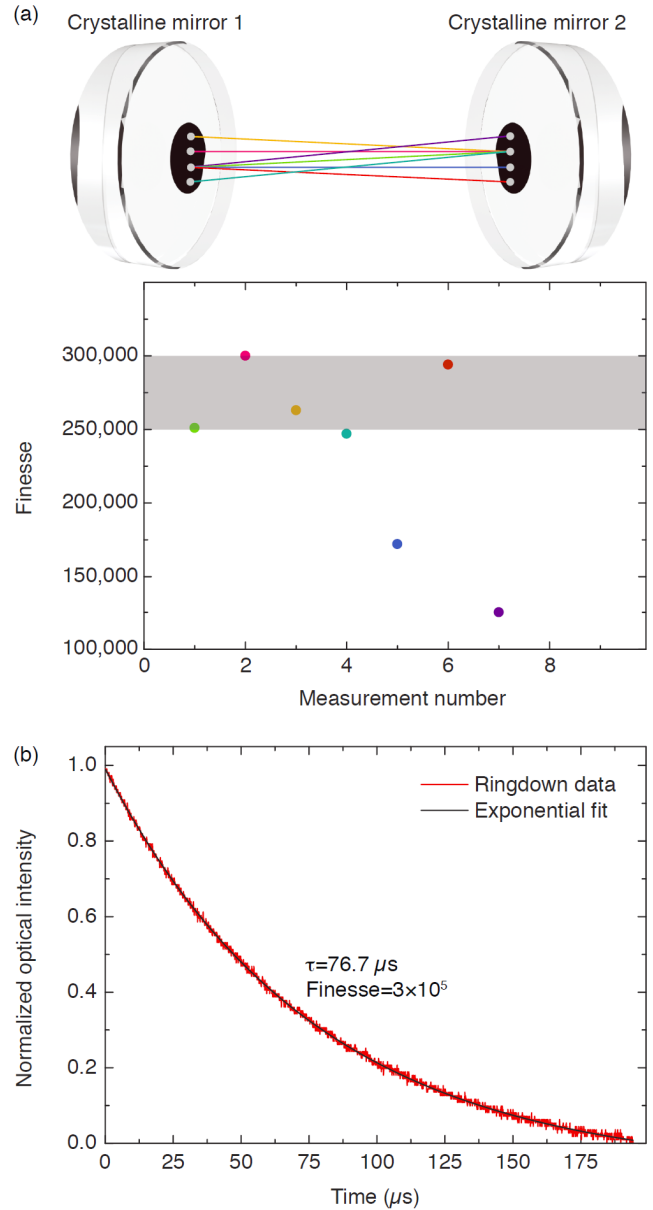

*Fig. 3.* Ultralow optical loss NIR crystalline coatings. a) Position-dependent finesse measurements for a 1550 nm coating with a nominal transmission of 5 ppm and a cavity length of 240 mm. The inset shows the relative position of each measured spot on the mirror pair compared with the 8-mm diameter coating disc; each data point is color coded to match the positioning. The positions with lowest loss yield finesse values between $2.5 \times 10^5$ and $3 \times 10^5$, highlighted in the gray band in the plot. Two points with increased losses yield a reduced finesse of $\sim 1 \times 10^5$. b) Normalized and fitted optical ringdown for the lowest loss position (point 2 in plot a). For this measurement the 1/e decay time is 77 μs, yielding an optical finesse of $3 \times 10^5$.

One important point to raise is the fact that our crystalline coatings show relatively strong birefringence when compared with competing amorphous reflectors. This has been explored in detail previously (see the Supplementary Information [17]). In our initial work with a 35-mm long cavity, we observed two orthogonal polarization eigenmodes separated by 4(0.4) MHz. We have now re-examined this with a 25-mm long cavity (linewidth of 2 kHz) and measure a splitting of 220 kHz. From these measurements we extract a cavity birefringence, θ, of ~1-5×10$^{-3}$. Contrast this with θ values in the 10$^{-4}$ to 10$^{-6}$ range that is typically observed in high finesse cavities employing amorphous interference coatings in the NIR [23-25] and MIR [26]. This is however not a detriment, just a unique feature to be aware of. For example, a strongly resolved polarization mode splitting, well beyond the cavity linewidth, can help optimize cavity coupling and minimize undesired optical feedback effects. It is also important to note that the birefringence is well controlled in these cavities and each coating disc incorporates a small flat, as can be seen in Fig. 1, which indicates the orientation of the crystal and thus the slow and fast axes of the coating.

In addition to the cavity ringdown tests, direct measurements of the optical absorption of these coatings have been carried out at a center wavelength of 1064 nm using the PCI method [21]. These measurements allow us to separate out the relative contributions of the loss components, something that is extremely important when optimizing the ultimate mirror performance. In this case we employ a 35.5-period GaAs/Al$_{0.92}$Ga$_{0.08}$As multilayer transferred to fused silica. The sample additionally includes a small Cr thin-film absorber for direct on-sample calibration.

Absorption measurements are realized using the PCI method with a reflected probe setup. In this case the probe is a low-noise 633 nm HeNe laser that allows for near shot-noise limited operation. One drawback of this configuration is the fact that the red probe light is heavily absorbed by the coating, which poses an obvious problem of additional carriers and ultimately absorption being generated in the measurement. Thus, care must be taken to extract the intrinsic coating absorption value by extrapolating back to an effective zero probe power (see Fig. 4 for an example). On the other hand, the use of a strongly attenuating probe helps to avoid unwanted interference effects within the coating, allowing for easier calibration of the absorption signal. Modeling shows that the single reflection from the air-coating interface dominates the PCI signal as a consequence of the relatively strong absorption of the above-bandgap HeNe beam in the high index GaAs surface layer.

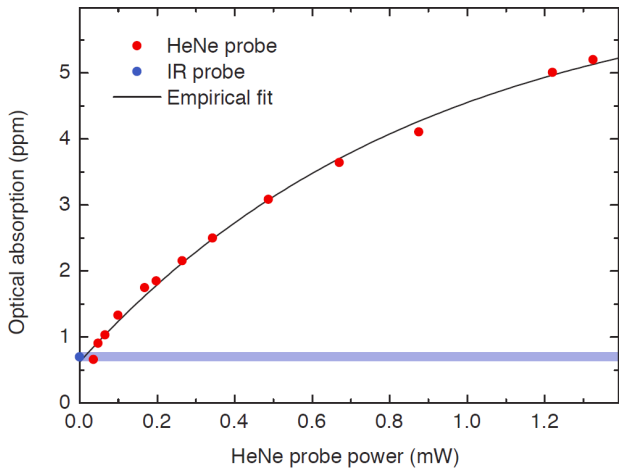

*Fig. 4. PCI measurement of sub-ppm optical absorption at 1064 nm in a 35.5-period substrate-transferred crystalline coating. The use of a visible (633 nm) HeNe probe laser in this system induces excess absorption via photogenerated carriers, these points are shown in red. With no probe-induced free-carrier losses, the absorption would remain independent of the applied probe power. The intrinsic absorption of the multilayer in the NIR is found by plotting this induced absorption as a function of the HeNe probe power, with the limiting absorption level corresponding to an effective zero probe power. The fitting function is based on the Beer-Lambert law assuming a power (and thus photon number) dependent photogenerated carrier concentration. The blue point corresponds to an induced absorption experiment using an additional transparent (1155 nm) probe: by cycling 0.9 mW of HeNe probe power off and on, the transparent NIR probe detects a nearly 6-fold change in the absorption signal, thus confirming and quantifying the free carrier effect for the red probe. The noise floor for this measurements is 0.2 ppm. Both the transparent probe datum and the y-intercept of the absorbing HeNe probe yield consistent absorption values of 0.7 ppm and 0.62 ppm respectively in this coating at 1064 nm.*

In these tests, the absorption signal is experimentally calibrated with a metallic thin film absorber, a common approach when using PCI instruments. To enable improved accuracy in determining the correct calibration factor, we apply this absorber directly to the sample. Here, a portion of the substrate and crystalline coating surface are overcoated by a 5 nm-thick, semi-transparent evaporated Cr layer. This thin layer does not affect the shape of the thermal field but provides an easily measurable absorption reference that is integrated directly on the sample surface.

To address the excess loss imprinted on the NIR pump by probe-generated free carriers, the dependence of the measured absorption with varying probe power is explored (Fig. 4). These measurements are made with a 5 W CW 1064 nm laser, yielding a limiting sensitivity better than 0.2 ppm. From this test, we conclude that the coating exhibits significantly less than 1 ppm of optical absorption at 1064 nm at the zero probe-power limit, with a measured absorption near 0.7 ppm at the point where the probe is attenuated to the system noise floor. Verification of this low absorption value has additionally been made by swapping out the HeNe laser for a transparent 1155 nm NIR probe. Simultaneous measurements with both transmitted 1155 nm and reflected 633 nm probes verify sub-ppm absorption in this mirror.

Ultimately, PCI measurements at 1064 nm and room temperature show that coatings generated with an optimized crystal growth process are capable of absorption values well below 1 ppm. Undertaking linear and 2-D scans of the mirror surface, we do however see potential for point-like high absorption regions in the films that require further investigation. These may be related to as yet unidentified defects in the epitaxial structure, or may simply be caused by debris on the mirror surface. Regardless, the current background loss level appears to be a consequence of free carrier absorption in the semiconductor multilayer and thus scales with the impurity-generated doping in the thin films. This assumption is further supported by growth tests that reveal an indirect dependence between the achievable absorption and the multilayer deposition rate. Thus it appears that a constant flux of impurities from the growth chamber are minimized by depositing the multilayer as fast as possible, while of course maintaining a reasonable surface quality and thickness control. Further reduction in the absorption may be realized by turning off all excess heat sources in the MBE system, providing further evidence for a background-limited absorption mechanism. Thus, for further reduction in the limiting optical absorption, it is imperative that the crystal growth system has as low of a background impurity level as possible, ultimately leading to a lower density of mobile charge carriers in the crystalline coatings. Future temperature-dependent optical absorption measurements should elucidate the microscopic details of this process.

Given the low background doping and thus average absorption in our crystalline coatings, coupled with the tight thickness control corresponding to sub-ppm transmission deviations, it appears that the key limitation to realizing further improvements in our NIR reflectors is predominantly optical scatter. Micro-roughness measurements by atomic force microscopy yield RMS values of 1.2 Å for our typical coatings. A simple surface-only estimate [27], valid in the long wavelength (for surface features much smaller than the wavelength) and high reflectivity limit [28], yields a corresponding scatter loss value of 2 ppm at 1064 nm. The additional few ppm of excess loss that we currently observe shows that we are still limited by extraneous scatterers in the multilayer, likely a combination of oval defects, surface contamination, and debris. Future efforts will focus on minimizing the occurrence of these loss centers by using a dedicated ultra-clean MBE system (equipped with the latest generation of Ga cells) and using optimized growth conditions and substrate preparation techniques to achieve ultra-smooth and low defect density epitaxial films.

In summary, through refinements in the deposition and microfabrication process, we have now reached a point where we can confidently exploit the intrinsically high optomechanical quality of our single-crystal multilayers. Inherent to the high purity and near perfect order of the constituent materials, these structures exhibit ppm-levels of optical losses while simultaneously exhibiting minimal mechanical dissipation. Furthermore, though not explicitly investigated here, these coatings exhibit significantly higher thermal conductivity than traditional amorphous multilayers. This unique combination of materials properties makes crystalline coatings ideal candidates for use in precision interferometry applications where coating Brownian noise currently represents a significant hurdle to further performance improvements [9]. At the same time, the ability to realize low optical losses with excellent thermal properties opens up unique application areas in high-power laser systems. Future optimization efforts will focus on the reduction of optical-scatter limiting defects in the initial epitaxial growth process, on optimization of the substrate and etch-stop removal, as well as on the final cleaning step in order to further reduce the impact of defect and particulate-limited optical scattering, with the aim of pushing the maximum attainable finesse to $5\times10^5$ and beyond.

## 3. Mid-infrared crystalline coatings

The implementation of our crystalline coating technology is not limited solely to operation in the NIR and can in fact be extended to longer wavelengths into the MIR region of the electromagnetic spectrum. Many large molecules of interest for atmospheric science, medicine, and national security have fundamental vibrational transitions in this range, making it ideal for trace gas detection efforts. However, it can be challenging to transition well-developed NIR technologies and techniques to the MIR, as many materials become lossy or even opaque in this range. As a result, technological progress in the MIR lags significantly behind its shorter-wavelength counterpart.

Cavity-enhanced direct frequency comb spectroscopy (CE-DFCS) has recently emerged as a powerful and versatile tool for sensitive and multiplexed spectroscopy in the MIR spectral region, having first been demonstrated in the NIR [29]. To reach high detection sensitivity, a frequency comb (exhibiting a roughly 100 nm spectral bandwidth) is injected into a high-finesse optical cavity to create approximately 100,000 simultaneous cavity-enhanced detection modes [30, 31]. The wealth of spectral information extracted from this measurement is processed in parallel, allowing for the generation of a complete molecular spectrum without scanning the laser source or employing wavelength-sensitive detection. This approach enables the unprecedented achievement of high spectral resolution, broad spectral coverage, and ultrasensitive detection of multiple molecular species simultaneously.

The versatility of the MIR CE-DFCS technique has been demonstrated in recent years through a number of experimental efforts. In 2013, the technique was used to probe trace (sub-ppm) amounts of hydrogen peroxide, $H_2O_2$, in percent-level concentrations of water [31]. This study relied on the broad bandwidth and high sensitivity of the CE-DFCS technique to demonstrate its utility for spectroscopic medical applications such as breath analysis, where interference from water-induced vibrational bands in the MIR can be a significant limitation. In 2014, the technique was utilized to probe short-lived chemical radicals, where the absorption enhancement provided by a MIR optical cavity was critical to the detection of trace concentrations of the highly-reacting species [32]. In this study, the transmitted cavity modes were spatially dispersed and recorded on a fast MIR InSb camera. In 2015, the CE-DFCS technique was integrated with helium buffer gas cooling technology to perform high-resolution spectroscopy on large rotationally-resolved molecules [33]. In this study, the high sensitivity, broad bandwidth, and high spectral resolution inherent in the CE-DFCS technique were again critical to the successful realization of the scientific objectives.

For all of its success, the CE-DFCS technique does have one large drawback. Due to its broad bandwidth, the optical frequency comb typically has a small power per comb tooth, making the CE-DFCS technique very sensitive to further losses in power. The most significant loss mechanism is currently due to excess optical losses, particularly scatter and absorption, in the mirrors comprising the optical enhancement cavity. The on-resonance transmitted power reduces quickly if the losses are comparable to the transmission [34]. In the MIR spectral region, established deposition techniques have yet to replicate their successes in the NIR spectral regions. Losses of up to 630 ppm in the MIR have been reported by Foltynowicz et. al [31] for high finesse cavity mirrors with a transmission of 203 ppm. In all of the experimental efforts described above, maintaining mirror reflectivity while lowering mirror loss will directly result in higher on-resonance transmission and higher absorption sensitivity. With even lower mirror losses, Doppler-free spectroscopic methods become accessible at room temperature for many common atmospheric molecules, such as methane [35].

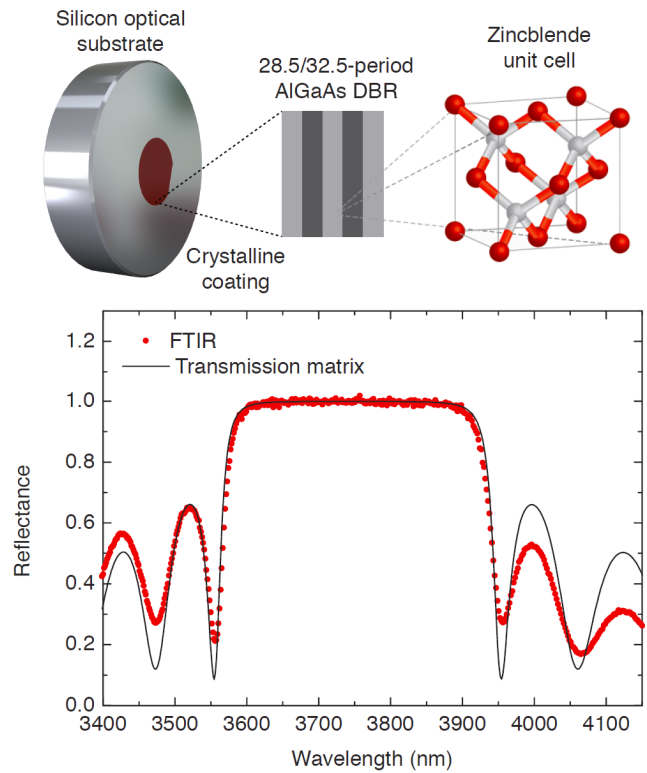

*Fig. 5. Prototype MIR crystalline coatings on silicon. Top: Basic schematic of the MIR reflectors characterized in this effort. The crystalline multilayer consists of either 28.5 or 32.5 periods of alternating high index GaAs layers and low index $Al_{0.92}Ga_{0.08}As$ layers transferred to super-polished single-crystal Si substrates with a 1-m radius of curvature and a broadband (3-4 μm) backside anti-reflection coating. Bottom: Reflectance spectrum measured via Fourier transform infrared spectroscopy (FTIR) for a 28.5-period coating centered at 3.7 μm. The stopband of this mirror spans nearly 300 nm making crystalline coatings promising reflectors for CE-DFCS.*

Here we present results for the first crystalline coatings designed for operation at 3.3 μm and 3.7 μm. Both mirror designs aimed at and achieved transmission and optical losses in the hundred ppm-range that were used to build enhancement cavities with a finesse near 10,000 at 3.3 μm and around 5,000 at 3.7 μm. The low losses of these mirrors allowed for cavity transmission values exceeding 21% and 47% of the coupled power at 3.3 μm and 3.7 μm, respectively.

The fabrication process for these mirrors is essentially identical to that of the NIR mirrors, with the multilayers consisting of stacks of alternating quarter-wave GaAs/$Al_{0.92}Ga_{0.08}As$ thin films grown by MBE. For the 3.3 μm structure the mirror consists of 32.5 periods of alternating 248.8-nm thick high index GaAs layers and 284.6-nm thick low index $Al_{0.92}Ga_{0.08}As$ (as-designed thickness of 17.3 μm for the coating). This multilayer was designed to have a nominal transmission of 227 ppm at 3.300 μm, while the 3.7 μm mirror employs 28.5 repeats of the high and low index layers respectively (281.1 nm of high index GaAs, 321.5 nm of low index $Al_{0.92}Ga_{0.08}As$, total thickness of 17.2 μm), aiming for a nominal transmission of 684 ppm at 3.725 μm. One potential difficulty encountered here is the significant thickness of the Bragg stack, which increases the overall defect density on the coating surface. This process is simply related to the required growth time (nearly 24 hours for our MIR mirror structures) and hence an accumulation of defects due to Ga spitting and particulate-driven contamination. In addition, we observe a crosshatched surface morphology, characterized by periodic surface undulations arising from strain relaxation due to dislocation glide in the film interior that is also associated with misfit dislocation formation at the substrate-epi-layer interface [36]. All of these processes are thickness-driven and thus become more problematic for reflectors with a longer center wavelength. The accumulation of these defects ultimately yields an excessively rough surface, precluding us from implementing the "flipped" bonding process for these initial MIR mirror prototypes. The results shown here are largely limited by scatter and thus currently represent conservative upper limits to the optical losses of these devices.

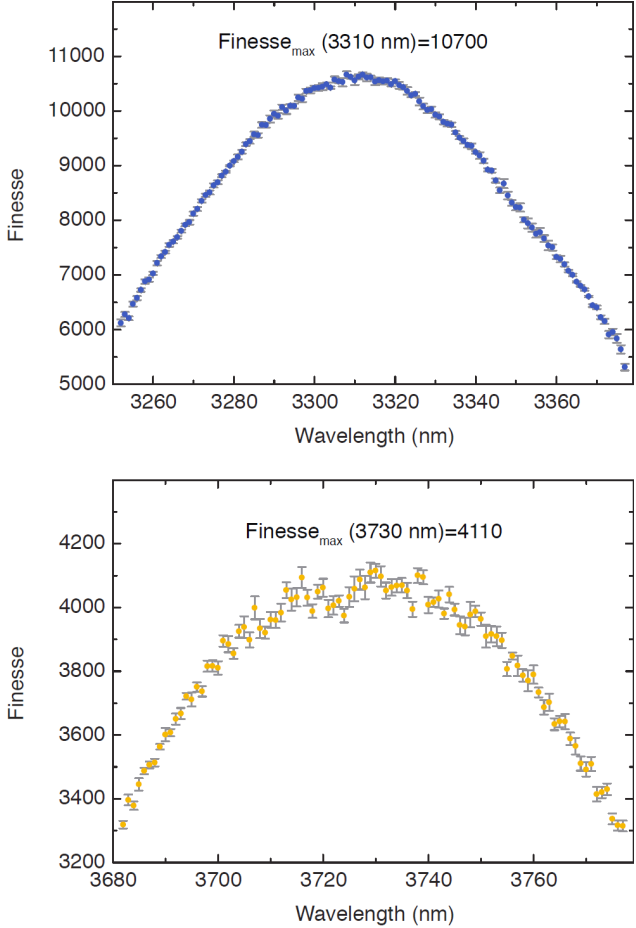

*Fig. 6. Finesse measurements for the 3.3 μm (top) and 3.7 μm (bottom) mirror pairs. Each cavity, of length 0.5488 m, was excited with a broad-bandwidth optical frequency comb and the transmitted light from the cavity was spectrally resolved using a reflection grating monochromator. At each wavelength, the ringdown time of the cavity was used along with its length to calculate the finesse. The error bars on both traces represent the +/- 1σ standard error from the measurement of 30 (top) and 10 (bottom) ringdown traces at each wavelength.*

Two prototype MIR mirror pairs, fabricated with our traditional non-flipped or right-side-up transfer process and designed with center wavelengths of 3.3 μm and 3.7 μm, respectively, were used to construct optical enhancement cavities. The ringdown time and reflection contrast were measured as a function of wavelength for each cavity. This is sufficient to partition total mirror losses (L+T) into transmission (T) and absorption+scatter (L) [34]. For these tests, we couple a broadband (~200 nm spectral width) MIR optical frequency comb [37] into the optical cavity. We ensure that the coupled light is spatially single mode by transmitting it through a single mode optical fiber prior to the cavity. The cavity FSR of 272 MHz is tuned to be double the repetition rate of the comb. This ensures that every other frequency comb mode is matched with a cavity mode [38]. In order to measure the ringdown time of the cavity, its transmitted beam is coupled into a second single-mode optical fiber before a portion of the MIR spectrum is selected using a calibrated reflection grating monochromator for analysis using a fast (10 MHz) MIR optical diode. The optical cavity finesse is recorded as a function of wavelength by tuning the center wavelength of the monochromator and measuring the decay rate of the transmission.

Measurement of the reflection contrast is realized by installing a beam splitter directly preceding the cavity. The rejected light is routed into the same detection path as the transmitted light. The cavity length is swept very slowly in this case to achieve maximum buildup and the ratio of the on-resonance and off-resonance cavity reflection is recorded as a function of wavelength.

The results of the finesse measurements are summarized in Table 1, with select data included in Figure 6. Our measurements yield maximum cavity finesse values of 10660 and 4100 for the 3.3 μm and 3.7 μm mirror pairs, respectively. Most importantly, the reflection contrast of the mirror pairs was measured to be 71% and 90% at their respective center wavelength. These two measurements yield average mirror losses of 159 ppm at 3.3 μm and 242 ppm at 3.7 μm with on-resonance cavity transmissions of 21% and 47% of the coupled power.

*Table 1: Measured transmission ($T_{measured}$), losses ($L_{measured}$), Finesse and cavity transmission ($T_{cavity}$) at the design wavelength (λ) of the mirrors. $L_{measured}$ and $T_{measured}$ are calculated from the measured Finesse and on-resonance cavity transmission, $T_{cavity}$. The design transmission, $T_{design}$, is also included for comparison.*

| λ [nm] | $T_{design}$, $T_{measured}$ [ppm] | $L_{measured}$ [ppm] | Finesse | $T_{cavity}$ |
|---|---|---|---|---|
| 3300 | 227, 136 | 159 | 10,660 | 21% |
| 3725 | 684, 524 | 242 | 4,100 | 47% |

Future efforts will focus on direct absorption measurements for these multilayers in order to de-convolve the relative contributions of optical absorption and scatter from the overall excess loss values. Simulations of the ultimate limits of optical absorption suggest a promising path forward for the development of high-performance AlGaAs-based MIR mirrors. In Figure 7 we present the results of a theoretical calculation for the limiting absorption of unintentionally doped (n-type background at $10^{14}$ cm$^{-3}$) GaAs/Al$_{0.92}$Ga$_{0.08}$As Bragg mirrors for center wavelengths between 850 and 5500 nm. This plot incorporates the wavelength dependence of the material absorption coefficient for GaAs [39] scaled to the average loss of the multilayer, the dispersion of both the high index GaAs layers, as well as the low index Al$_{0.92}$Ga$_{0.08}$As films [40], and additionally includes the variation in penetration depth with wavelength for the Bragg mirror [41]. Assuming a similar wavelength dependence for the absorption coefficient of the high-Al content layers, absorption losses significantly below 100 ppm can be achieved from 900 to roughly 7500 nm.

Extending the maximum operating wavelength, while maintaining such low loss levels, will require implementation of an ultra-pure and low defect density crystal growth system and process. Longer term research may additionally involve the exploration of alternative materials, particularly in the case of the low index films, where a reduction in the refractive index may both increase the mirror bandwidth and significantly reduce the multilayer thickness for a given transmission level. One interesting example is the demonstration of epitaxial Bragg mirrors based on GaAs/Ba$_x$Ca$_{1-x}$F$_2$ [42]. Recent work on single-crystal fluoride-based whispering-gallery mode resonators has shown the potential for ultralow optical losses in such materials for wavelengths out to 5 μm [43]. Even in the absence of background contamination in the epitaxial growth process, the ultimate long wavelength limit will be set by optical phonon absorption in the crystal structure, which leads to significant losses in the so-called reststrahlen band beyond about 20 μm in GaAs [44]. Regardless, from these initial investigations, we are confident that these unique coatings represent a promising solution for the realization of ultra-low loss MIR reflectors.

The loss values of our prototype MIR mirrors are already extremely promising, with excess loss levels comparable to state-of-the-art MIR coatings in this wavelength range. For future development efforts we plan to follow a similar route as for the NIR mirrors where we have now demonstrated excess loss values, including both scatter and absorption, below 5 ppm. Based on calculations of the limiting optical losses in the mid-wave regime, we anticipate significantly improved optical performance in future generations of MIR-focused crystalline coatings.

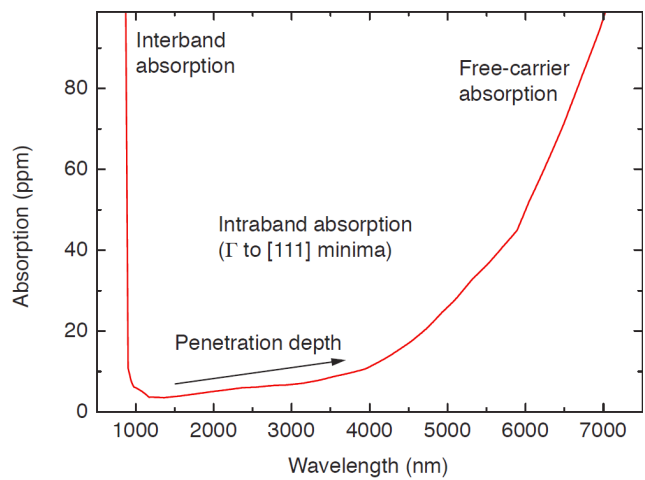

*Fig. 7. Calculated absorption loss as a function of center wavelength for unintentionally-doped GaAs/Al$_{0.92}$Ga$_{0.08}$As Bragg mirrors. This theoretical curve assumes background n-type doping in the epitaxial films with a concentration at the low $10^{14}$ cm$^{-3}$ level, yielding ppm-level losses in the near IR (from 1000-1600 nm) as verified experimentally. The curve also takes into account the dispersion (i.e. the variation in refractive index with frequency) of both the high and low index layers, as well as the variation in penetration depth with wavelength for mirrors covering the range from just below the GaAs absorption edge at ~870 nm at 300 K out to 7 μm, where free-carrier losses become a significant impediment. The kinks in the dataset are a consequence of the interpolation function used to generate the final absorption curve.*

## 4. Conclusions

The key advantages of our newly developed crystalline coatings over state-of-the-art IBS-deposited dielectric multilayers are: (1) drastically reduced Brownian noise and (2) a greater than 30-fold enhancement in thermal conductivity (~1 Wm$^{-1}$K$^{-1}$ for IBS films compared with ~30 W/m$^{-1}$K$^{-1}$ for GaAs/AlGaAs multilayers [10]). While these advantages have been demonstrated previously [17, 19], it has remained an open question as to whether their ultimate optical performance is—in theory and practice—competitive with advanced dielectric coatings. Here we have methodically investigated the optical properties of these novel coatings in the NIR and MIR and show conclusively that this technology is capable of comparable levels of optical performance, rivaling that of modern IBS-deposited multilayers in the NIR. Work is already underway towards the realization of new benchmarks in laser frequency stability through the use of ultrastable optical reference cavities incorporating end mirrors with crystalline coatings. At the same time, crystalline coating enabled MIR enhancement cavities are being used for frequency comb spectroscopy experiments in order to study reaction kinetics and complex molecules. Thus, crystalline coatings exhibit substantially reduced Brownian noise, excellent thermal conductivity, and potentially the widest spectral coverage of any single supermirror technology.

Through a reduction in the background doping via an optimized MBE process, coupled with a modified microfabrication process, we demonstrate NIR reflectors with excess optical losses as low as 3 ppm, enabling the realization of 1550 nm cavity end mirrors yielding a finesse of $3\times10^5$ (for a coating transmission of 5 ppm and demonstrated for cavity lengths up to 24 cm). Moreover, for the first time in the same coating system, we can simultaneously realize low optical losses in the MIR. As a first demonstration, we have fabricated and characterized prototype 3.3 and 3.7 μm crystalline coatings on single-crystal silicon substrates with excess losses at the hundred ppm level. With further refinements to this technology, we project that it would be possible to maintain excess losses well below 100 ppm in these coatings to center wavelengths of roughly 7.5 μm. It is important to note that crystalline coatings do not exhibit discrete absorption bands over this range, for example lacking the OH-mediated "water" absorption window that is found in IBS-derived metal oxide coatings near 1400 nm. This unique combination of simultaneous low losses in the near and mid-IR gives us the freedom to develop novel optics including low-loss NIR/MIR dichroic reflectors. Crystalline coatings can ultimately be implemented in a wide variety of demanding applications including but not limited to the construction of ultrastable optical cavities exhibiting minimal Brownian noise for advanced metrology applications, high-finesse MIR reflectors for cavity-ringdown based optical gas sensing, and low-loss reflectors capable of efficient heat transport for industrial laser-based manufacturing systems.

**Funding**. CMS: Austria Wirtschaftsservice Gesellschaft (aws; Seed Financing P130811 7-SZIO1), European Research Council (ERC; Proof of Concept grant 310736), European Association of National Metrology Institutes and European metrology research program (EURAMET/EMRP; QESOCAS, EXL01-REG4), Defense Advanced Research Projects Agency (DARPA; FAA-9550-14-C-0030); Vienna: Austrian Science Fund (FWF; I909); JILA: DARPA, AFOSR, NIST, and NSF Physics Frontier Center. O. Heckl is partially supported by a Humboldt Postdoctoral Fellowship and L. Sonderhouse is supported by an NSF Graduate Fellowship.

**Acknowledgment**. We thank Robert Yanka, Seokjae Chung, and Chris Santana from IQE for the growth of the epitaxial multilayers. The CMS team acknowledges the dedication of Dr. Christian Pawlu and thanks him for a number of insightful discussions as well as his input on this manuscript. CMS additionally thanks Markus Stana from the Nanostsrukturzentrum of the University of Vienna for his efforts in characterizing various defects in the crystalline coatings. A portion of this work was performed in the UCSB Nanofabrication Facility.